# Asymmetric scattering behaviors of spin wave dependent on magnetic vortex chirality

**Xue-Feng Zhang[1#], Je-Ho Shim[2#], Xiao-Ping Ma[2*], Cheng Song[3], Haiming Yu[4] and Hong-Guang Piao[1,2*]**

*1 Hubei Engineering Research Center of Weak Magnetic-Field Detection, China Three Gorges University, Yichang 443002, P. R. China*

*2 Department of Physics, College of Science, Yanbian University, Yanji 133002, P. R. China*

*3 Key Laboratory of Advanced Materials (MOE), School of Materials Science and Engineering, Tsinghua University, Beijing 100084, P. R. China*

*4 Fert Beijing Institute, School of Integrated Circuit Science and Engineering, Beijing Advanced Innovation Center for Big Data and Brain Computing, Beihang University, 100191 Beijing, P. R. China*

*** Corresponding author's E-mails:** xpma1222@ybu.edu.cn; hgpiao@ybu.edu.cn

**Abstract**

In this letter, an asymmetric spin wave scattering behaviors caused by vortex chirality are investigated in cross-shaped ferromagnetic system. In the system, four scattering behaviors are found, 1) asymmetric skew scattering, depending on the polarity of vortex core, 2) back scattering (reflection), depending on the vortex core stiffness, 3) side deflection scattering, depending on structural symmetry of the vortex circulation, and 4) geometrical scattering, depending on waveguide structure. The first and second scattering behaviors are attributed to nonlinear topological magnon spin Hall effect related to magnon spin-transfer torque effect, which has value for magnonic exploration and application.

Nonlinear science is an interdisciplinary science which studies the common law of nonlinear behaviors in various systems [1,2]. In recent years, the nonlinear interaction between spin wave (magnon) and topological magnetic texture (magnetic soliton) has received a lot of attention in magnetic systems [3,4], because it provides a good platform for exploring nonlinear science and has potential application prospects [5-7]. In many nonlinear magnetic systems, the equilibrium position (vortex core position in the steady state) of magnetic vortices exhibits notable robustness [8]. In other words, the vortex structure remains remarkably stable even under the interference of the external field. Hence, the interaction between spin waves (SWs) and magnetic vortices can provide a stable platform for the study of nonlinear dynamics and may have application prospects in the field of spintronics [9].

SWs, as a collective precession behavior of spins in the magnetic system [10,11], can transmit information using spin angular momentum rather than electron as a carrier, which can effectively solve the problem of thermal effect related power consumption caused by electron scattering. Since the SW has a wide frequency range from GHz to THz and its wavelength can also reach the nanometer scale [12,13], it is advantageous to transmit information in magnetic nanodevices. Particularly, since it can propagate in magnetic materials, wave properties like diffraction [14], refraction [15], and interference [16,17] can be used to realize a variety of useful functions. At present, some SW-based functional devices have been reported, such as SW-based diodes and logic devices [18-20]. Interestingly, a nonreciprocal behavior of SW propagation can be achieved by modulating Dzyaloshinskii-Moriya interaction [21,22] and magnetic anisotropy [23,24] in the magnetic system, which is expected to be applied to the magnonic logics as an asymmetric behavior. However, it is obvious that the best way to apply magnonics is not to control the SW behaviors by changing the inherent structural properties of magnetic materials. Therefore, it will be of great significance for the development of ultrafast, low power consumption and stable performance magnonic devices if the control of SW behavior can be realized through the reconstruction of magnetic texture [25].

Recently, it has been found that the behaviors of SW propagation become richer and easier to be controlled with the introduction of topological spin textures [4]. It is because that the spin texture exhibits structurally stable soliton-like dynamic behavior [26,27], which can produce SWs when excited by an external field and influences the behavior of SWs as they propagate [27–29]. Among various spin textures, the magnetic vortex not only effectively excites SWs [30,31] through the gyromotion [32,33], azimuth mode motion [34,35] or radial mode motion [36] of its core, but also controls the SW propagation behaviors by switching its chirality [30]. The vortex has two chirality with right and left handedness [37], which are characterized by the circulation sense of the in-plane circling spins ( $c = -1$ for clockwise (CW), $c = 1$ for counterclockwise (CCW)) and the polarity of the VC ($p = 1$ for upward core, $p = -1$ for downward core). That is, there are four chiral states: the vortex(1,1) and (-1,-1) for right-handed chirality, and vortex(1,-1) and (-1,1) for left-handed chirality. Therefore, if the propagation behavior of SW can be effectively manipulated by controlling the four chiral states of the magnetic vortex, it will benefit the design of magnonic devices and promote the exploration of nonlinear science.

In this paper, an asymmetric SW scattering behavior, depending on the chirality of magnetic vortices, is investigated in a cross-shaped ferromagnetic nanostructure. The results show that there are four scattering mechanisms in the magnetic system, including asymmetric skew scattering, back scattering (reflection), side deflection scattering, and geometrical scattering, which depend on the VC polarity, the VC stiffness, structural symmetry of the vortex circulation, and the waveguide structure, respectively. Moreover, based on the asymmetric effect of chiral topological structure on scattering behaviors of SWs, a control scheme of SW propagating in the cross-shaped nanostructure is proposed by reconstructing the chiral structure.

In order to investigate the effect of magnetic vortex chirality on the SW scattering behavior, the SW propagation in a cross-shaped ferromagnetic nanostructure was carefully explored by using micromagnetic simulations based on the Landau-Lifshitz-Gilbert equation [38]. In all the simulations, the material parameters of the Permalloy were considered, the saturation magnetization $M_S = 8.6 \times 10^5$ A/m, the exchange stiffness coefficient $A_{ex} = 13 \times 10^{-12}$ J/m, the Gilbert damping constant $\alpha = 0.01$, and the zero magnetocrystalline anisotropy. As schematically illustrated in Fig. 1, the cross-shaped nanostructure is composed of two flat rectangular strips of $2000 \times 200 \times 2$ nm$^3$, overlapping each other, which is discretized into $10^6$ unit cells of $2 \times 2 \times 2$ nm$^3$. Initially, a magnetic vortex is set in the cross region of the nanostructure, as shown in two insets of Fig. 1, which illustrates two chiral structures of vortex(1, 1) and vortex(1, -1). All spins in the four arms in the nanostructure are initially set to point toward the cross center (see the white arrows) to form a stable magnetic vortex structure. A SW source is placed at the end of the left arm (arm-1), as the black part shown in Fig.1. The SW is excited by the sinusoidal microwave field $B_{ext} = B_0 \sin(2\pi f t)\hat{z}$, with $B_0 = 200\,\mathrm{mT}$ and $f = 50\,\mathrm{GHz}$. An original monochromatic SW ($\mathrm{SW}_{org}$) propagates along the +$x$-axis direction from arm-1 to arm-3. To prevent the influence of SWs reflection from the ends, the high damping constant is set in the 30-nm-width region (see gray parts) at the three ends of each arm (arm-2, 3 and 4).

As shown in Fig. 2a, it can be clearly observed that when the $\mathrm{SW}_{org}$ propagating to the cross region of the nanostructure, the majority of the $\mathrm{SW}_{org}$ travels through the vortex structure, and reaches arm-3, while a minority of $\mathrm{SW}_{org}$ is scattered to arm-2 and arm-4. As shown in Fig. 2b, there is a significant difference (asymmetry) in the average amplitudes of the SWs scattered into arm-2 and arm-4, whereas no such significant difference is observed in the no-vortex case (see Fig. S1 in Supplemental Material). Hence, it can be found that there exists an interaction between the SW and the vortex, which suggests the occurrence of asymmetric behaviors in SW propagation. In addition, it is worth mentioning that in the vortex(1,1) case (see Fig. 2b), the average amplitude of SWs injected into arm-2 is about 12 times larger than that value injected into arm-4.

The asymmetric scattering of SWs was compared in the cases of four different chiral vortices (vortex(1,1), vortex(-1,1), vortex(-1,-1), and vortex(1,-1)) and the case without vortex (no-vortex) in order to verify the universality of the interaction between SW and vortex.. As shown in Fig. 2c, the percentages of the average amplitudes of SWs scattered into arm-2 (red) and arm-4 (green) were compared in each case. In the no-vortex cases, the percentages of SW amplitudes for arms 2 and 4 are nearly identical, whereas in the presence of vortex structures, the percentages of SW amplitudes for arms 2 and 4 are noticeably different. It is obvious that the asymmetric scattering behavior of SWs in this system occurs due to the interaction between SW and vortex. It is worth reminding that the SWs scattered into arm-2 are relatively dominant when the VC polarity is up ($p = 1$), while the SWs scattered into arm-4 are dominant when the VC polarity is down ($p = -1$). The change of vortex circulation ($c = \pm 1$) does not significantly affect the asymmetric scattering effect. All of these reveal that the asymmetric scattering effect of the SWs in this system is mainly dependent on the VC polarity rather than the vortex circulation.

In order to further confirm the dominant role of VC polarity on the asymmetric scattering behavior of SWs in this system, a no-VC case was simulated in the redesigned nanostructure with an $8 \times 8 \ \mathrm{nm}^2$ hole at the center of the cross region, as shown in Fig. 3a. The hole can pin the equilibrium point of the vortex circulation at the center of the cross region, and the vortex circulation structure still maintains symmetry (see inset of Fig. 3a). The results show that the asymmetric scattering behavior of SWs cannot be observed in the no-VC case, as shown in Fig. 3b. However, by comparing the no-VC case and the no-vortex case (see Figs. 3b and 3c), it can be clearly observed that the SW scattering effect in the no-VC case is more significant ($\sim 4.6$ times) than that in the no-vortex case, and it is found that there is a phase difference ($\Delta t \approx 3$ ps) between them, which suggests that the circulation structure of the magnetic vortex also has a scattering effect on the $\mathrm{SW}_{org}$. In the no-vortex case, it can be clearly observed that the SWs ($\mathrm{SW}_{sml}$) are symmetrically scattered into arm-2 and arm-4, as illustrated in Fig. S1 of Supplemental Material. This is called geometric scattering since the $\mathrm{SW}_{org}$ is attributed to a soliton mass loss effect caused by the geometric characteristics of the waveguide structure. Therefore, it can be considered that the source of $\mathrm{SW}_{sml}$ is located at the edges of both side of the cross region. Obviously, in the no-VC case, the SWs scattered into the arm-2 and arm-4 consist of two parts: one part is the $\mathrm{SW}_{sml}$ and the other part is the SWs caused by the vortex circulation scattering ($\mathrm{SW}_{C}$).

The asymmetric scattering behavior caused by the interaction between SW and VC is an interesting physical phenomenon which belongs to the field of nonlinear science. In order to explore this phenomenon, the vortex(1,1) case was carefully analyzed below. In Fig. 4a, it can be clearly observed that 600 ns after the VC is excited by the $\mathrm{SW}_{org}$, its gyromotion is effectively suppressed under the mechanical action of damping ($F_{DM}$, red arrows), and finally converges (see the red time region) at a position ~1.24 nm away from the cross center. This means that the $\mathrm{SW}_{org}$ also has a skew mechanical effect on the VC, that is, a SW current force [39]. Although the VC deviates from the cross center under the $\mathrm{SW}_{org}$ current force ($F_{SW}$, the green arrow), it will not deviate further after ~1.24 nm shift because there is a restoring force ($F_{EP}$, the blue arrow) in this system that always points the cross center and its magnitude is related to the distribution of stray fields around the VC. Therefore, in this system the direction of the $F_{EP}$ is always opposite to the $F_{SW}$ while its magnitude is always equal to the $F_{SW}$ and depends on the distance (R) of the equilibrium position (EP) of the vortices from the cross geometric center. That is, the EP is determinde by $F_{EP}$ and $F_{SW}$. Nevertheless, after the equilibrium position of the vortex shifts a certain distance under the combined action of $F_{EP}$ and $F_{SW}$, the gyromotion behavior of the VC still follows the Thiele equation [39,40], as shown by the dotted fitting line in Fig. 4a. The Thiele equation is written as bellow,

$$-\mathrm{G} \times \dot{R} = F_{DM} + F_{EP} + F_{SW} \qquad (1),$$

where, $G$ is gyro vector of VC (G > 0, if $p = 1$). The $F_{DM} = D \cdot \dot{R}$, and the direction always point to the EP, $D$ is damping tensor. The $F_{EP} = -k_{VC} R$, and its direction is opposite to the direction of the external force, and always points to the geometric center of the cross region, $k_{VC}$ is stiffness coefficient. The external force, $F_{SW} = F_0 \left( \sigma_{\parallel} \hat{k}_x + \sigma_{\perp} (\hat{z} \times \hat{k}_x) \right)$, which direction is opposite to the $F_{EP}$. Here, $F_0$ strength depends on the $\mathrm{SW}_{org}$ amplitude and the Gilbert damping. The $F_{SW}$ acting on the VC, can be resolved into two parts, that is, $F_{SW}^{\parallel}$ and $F_{SW}^{\perp}$, which are determined by $\sigma_{\parallel}$ ($\sigma \parallel \hat{x}$) and $\sigma_{\perp}$ ($\sigma \perp \hat{x}$), respectively. The $\sigma_{\parallel}$ and $\sigma_{\perp}$ are determined by the skew scattering angle $\theta$ of the $\mathrm{SW}_{org}$ and the differential of cross-section to $\theta$, $\mathrm{d}\sigma/\mathrm{d}\theta$ [41,42]. The $F_{SW}^{\parallel}$ and $F_{SW}^{\perp}$ can be calculated according to the EP position $(F_0 \sigma_{\parallel}/k_{VC},\ -F_0 \sigma_{\perp}/k_{VC})$, *e.g.*, the EP for the vortex(1,1) case is at (0.89 nm, -0.88 nm). As shown in Fig. S2 of Supplementary Material, the VC will eventually converge to the EP in all vortex cases, and the EP depends on the $F_{SW}$ and VC polarity. For instance, for the cases of $p = -1$ ($p = 1$) VC, the EP will shift to the upper-right (lower-right) corner. This means that the EP shift is attributed to the skew scattering effect of the $\mathrm{SW}_{org}$, which results in the reflection and scattering of the $\mathrm{SW}_{org}$ itself, thus forming the asymmetric scattering behavior of the SW in the cross-shaped nanostructure.

In order to further confirm the possibility of SW scattering behavior caused by the reaction of $F_{SW}$ on the VC, the distribution of SWs around the VC in the no-vortex and vortex(1,1) cases are compared. In the signals detected in the arm-1, it is found that the mean amplitude of the signals in the vortex(1,1) case is obviously larger than that in the no-vortex case, $\Delta\overline{m}_z \approx 0.0025$, indicating that the VC reflects the $\mathrm{SW}_{org}$ in the vortex(1,1) case, as shown in Fig. 4b. In the signals detected in the arm-2, it is also found that the mean amplitude of the signals in the vortex(1,1) case is significantly larger than that in the no-vortex case, $\Delta\overline{m}_z \approx 0.0016$, which is attributed to the skew scattering of the $\mathrm{SW}_{org}$ in the vortex(1,1) case, as shown in Fig. 4c. However, in the signals detected in the arm-4, although the mean amplitude of the signals in the vortex(1,1) case is found to be larger than that in the no-vortex case, $\Delta\overline{m}_z \approx 0.0005$, it is obviously weaker than in the arm-2 ($\Delta\overline{m}_z \approx 0.0016$). This means that the skew scattering effect of the VC on the $\mathrm{SW}_{org}$ obviously has asymmetric nature, as shown in Fig. 4c. According to the above phenomena, the conclusions are as follows: 1) The VC can cause reflection or skew scattering of the SWs; 2) The reaction force $F_{SW}$ of the scattering or reflection will cause the oblique shift of vortex EP; 3) The skew scattering effect of VC on the SW is asymmetric and depends on the VC polarity (see Fig. S2 of Supplementary Material). At present, the asymmetric interaction between VC and SW is considered originating from a so-called nonlinear topological magnon spin Hall effect, which related to magnon spin-transfer torque effect [43-45], and depends on the magnon chirality and the VC polarity. However, the scattering behavior of vortex circulation structure on SWs has not received much attention so far, because the asymmetric scattering effect induced by vortex circulation structure is insignificant compared with that of the VC, especially in the case of the SW driving the VC motion. In this paper, it is found that the vortex circulation structure contributes to the asymmetric scattering behavior of the SW and cannot be ignored. The obliqued shift of VC will make the vortex circulation distribution in the cross region of the nanostructure asymmetric, which will form an asymmetric scattering cross-section for the propagating SW, and eventually lead to asymmetric scattering effect. This conclusion is further confirmed by artificially designing a new micromagnetic simulation in which the vortex EP in the cross region is not located at the geometric center (see Fig. 3a).

In conclusion, it is found that there are four scattering mechanisms for the SWs travelling in the cross-shaped ferromagnetic nonlinear system, as schematically illustrated in Fig. 5, as follows: 1) the asymmetric skew scattering (SW$_P$, purple arrow), related to VC polarity; 2) the reflection (SW$_{re}$, red curved arrow), associated with the scattering cross-section and stiffness of VC; 3) the side deflection scattering (SW$_C$, yellow arrows), depending on structural symmetry of vortex circulation; and 4) the geometrical scattering (SW$_{sml}$, green arrows), depending on waveguide structural characteristics. The four mechanisms will play an important role in the exploration of nonlinear science and the development of magnonic devices.

This work was supported by the Basic Science Research Program of the National

Research Foundation of Korea (No. 2021R1F1A1050539), the Yanbian University Research Project (No. 482022104), and the Yichang natural science research project (No. A22-3-010). Z.X.F. and S.J.H. contributed equally to this work.

## AUTHOR DECLARATIONS

**Conflict of Interest**: The authors have no conflicts to disclose.

## DATA AVAILABILITY

The data that support the findings of this study are available within the article and its supplementary material or from the corresponding author upon reasonable request.

## References

[1] P. G. Kevrekidis, J. C.-Maraver, A. Saxena, eds. New York: Springer (2020).

[2] O. Lee, K. Yamamoto, M. Umeda, C. W. Zollitsch, M. Elyasi, T. Kikkawa , E. Saitoh, G. E. W. Bauer, and H. Kurebayashi, Phys. Rev. Lett. **130**, 046703 (2023).

[3] A. Barman, G. Gubbiotti, et al., J. Phys-Condens. Mat. **33** 413001 (2021).

[4] H. Yu, J. Xiao, and H. Schultheiss, Phys. Rep. **905**, 1 (2021).

[5] F. G.-sanchez, P. Borys, R. Soucaille, J.-P. Adam, R.L. Stamps, J.-V. Kim, Phys. Rev. Lett. **114** 247206 (2015).

[6] Z. Wang, H. Y. Yuan, Y. Cao, and P. Yan, Phys. Rev. Lett. **129**, 107203 (2022).

[7] Z. Li, M. Ma, Z. Chen, K. Xie, and F. Ma, J. Appl. Phys. **132**, 210702 (2022).

[8] K. Y. Guslienko, B. A. Ivanov, V. Novosad, Y. Otani, H. Shima, and K. Fukamichi, J. Appl. Phys. **91**, 8037 (2002)

[9] H.-K. Park, J.-H. Lee, J. Yang, and S.-K. Kim, J. Appl. Phys. **127**, 183906 (2020).

[10] A. V. Chumak, V. I. Vasyuchka, A. A. Serga, and B. Hillebrands, Nature Phys. **11**, 453 (2015).

[11] V. V. Kruglyak, S. O. Demokritov, and D. Grundler, J. Phys. D: Appl. Phys. **43**, 264001 (2010).

[12] C. Liu, J. Chen, T. Liu, et al., Nat. Commun. **9**, 1 (2018).

[13] I. Razdolski, A. Alekhin, N. Ilin, J. P. Meyburg, V. Roddatis, D. Diesing, U. Bovensiepen and A. Melnikov, Nat. Commun. **8**, 1 (2017).

[14] S. Mansfeld, J. Topp, K. Martens, J. N. Toedt, W. Hansen, D. Heitmann, and S. Mendach, Phys. Rev. Lett. **108,** 047204 (2012).

[15] J.-N. Toedt, M. Mundkowski, D. Heitmann, S. Mendach and W. Hansen, Sci. Rep. **6**, 1 (2016).

[16] Á. Papp, W. Porod, Á. I. Csurgay, and G. Csaba, Sci. Rep. **7**, 1 (2017).

[17] M. Balinskiy, H. Chiang, D. Gutierrez, and A. Khitun, Appl. Phys. Lett. **118,** 242402 (2021).

[18] J. Lan, W. Yu, R. Wu, and J. Xiao, Phys. Rev. X **5**, 041049 (2015).

[19] K. Szulc, P. Graczyk, M. Mruczkiewicz, G. Gubbiotti, and M. Krawczyk, Phys. Rev. Appl. **14**, 034063 (2020).

[20] K.-S. Lee, and S.-K. Kim, J. Appl. Phys. **104**, 053909 (2008).

[21] J. Chen, H. Yu, and G. Gubbiotti, J. Phys. D: Appl. Phys. **55**, 123001 (2021).

[22] H. Wang, J. Chen, T. Liu, et al., Phys. Rev. Lett. **124**, 027203 (2020).

[23] O. Gladii, M. Haidar, Y. Henry, M. Kostylev, and M. Bailleul, Phys. Rev. B **93**, 054430 (2016).

[24] J. H. Kwon, J. Yoon, P. Deorani, J. M. Lee, J. Sinha, K.-J. Lee, M. Hayashi, and H. Yang, Sci. Adv. **2**, 1501892 (2016).

[25] E. Albisetti, D. Petti, G. Sala, et al. Commun. Phys. **1**, 56 (2018)

[26] A. F.-Pacheco, R. Streubel, O. Fruchart, R. Hertel, P. Fischer, and R. P. Cowburn, Nat. Commun.**8**, 1 (2017).

[27] X.-P. Ma, J. Zheng, H.-G. Piao, D.-H. Kim, and P. Fischer, Appl. Phys. Lett. **117**,062402 (2020).

[28] S. Mayr, L. Flajšman, S. Finizio, A. Hrabec, M. Weigand, J. Förster, H. Stoll, L. J. Heyderman, M. Urbánek, S. Wintz, and J. Raabe, Nano Lett. **21**, 1584 (2021).

[29] J. Chen, J. Hu, and H. Yu, ACS Nano **15**, 4372 (2021).

[30] S. Wintz, V. Tiberkevich, M. Weigand, J. Raabe, J. Lindner, A. Erbe, A. Slavin, and J. Fassbender, Nat. Nanotechnol. **11**, 948 (2016).

[31] L.-J. Chang, J. Chen, D. Qu, L.-Z. Tsai, Y.-F. Liu, M.-Y. Kao, J.-Z. Liang, T.-S. Wu, T.-M. Chuang, H. Yu, and S.-F. Lee, Nano Lett. **20**, 3140 (2020).

[32] M. Kammerer, M. Weigand, M. Curcic, et al., Nat. Commun. **2**, 279 (2011).

[33] X.-P. Ma, H. Yang, C. Li, C. Song, and H.-G. Piao, Chin. Phys. Lett. **38**, 127501 (2021).

[34] J. P. Park and P. A. Crowell, Phys. Rev. Lett. **95**, 167201 (2005).

[35] M. Sproll, M. Noske, H. Bauer, M. Kammerer, A. Gangwar, G. Dieterle, M. Weigand, H. Stoll, G. Woltersdorf, C. H. Back, and G. Schütz, Appl. Phys. Lett. **104**, 012409 (2014).

[36] X.-P. Ma, M.-X. Cai, P. Li, J.-H. Shim, H.-G. Piao, and D.-H. Kim, J. Magn. Magn. Mater. **502**, 166481 (2020).

[37] V. Uhlíř, M. Urbánek, L. Hladík, J. Spousta, M.-Y. Im, P. Fischer, N. Eibagi, J. J.

Kan, E. E. Fullerton and T. Šikola, Nat. Nanotech. **8**, 341 (2013).

[38] A. Vansteenkiste, J. Leliaert, M. Dvornik, M. Helsen, F. G.-Sanchez, and B. V. Waeyenberge, AIP Adv. **4,** 107133 (2014).

[39] S. Li, J. Xia, X. Zhang, M. Ezawa, W. Kang, X. Liu, Y. Zhou, and W. Zhao, Appl. Phys. Lett. **112**, 142404 (2018).

[40] J.-H. Shim, H.-G. Piao, S. H. Lee, S. K. Oh, S.-C. Yu, S. K. Han, and D.-H. Kim, Appl. Phys. Lett. **99**, 142505 (2011).

[41] C. Schütte and M. Garst, Phys. Rev. B **90**, 094423 (2014).

[42] C. Schütte, J. Iwasaki, A. Rosch, and N. Nagaosa, Phys. Rev. B **90**, 174434 (2014).

[43] M. Mochizuki, X. Z. Yu, S. Seki, N. Kanazawa, W. Koshibae, J. Zang, M. Mostovoy, Y. Tokura, and N. Nagaosa, Nat. Mater. **13**, 241 (2014).

[44] Z. Gao, F. Wang, X. Zhao, T. Wang, J. Hu, P. Yan, arXiv:2212.01172, (2023). https://doi.org/10.48550/arXiv.2212.01172

[45] Z. Jin, X. Yao, Z. Wang, H. Y. Yuan, Z. Zeng, Y. Cao, P. Yan, arXiv:2301.03211 (2023). https://doi.org/10.48550/arXiv.2301.03211

**Figures**

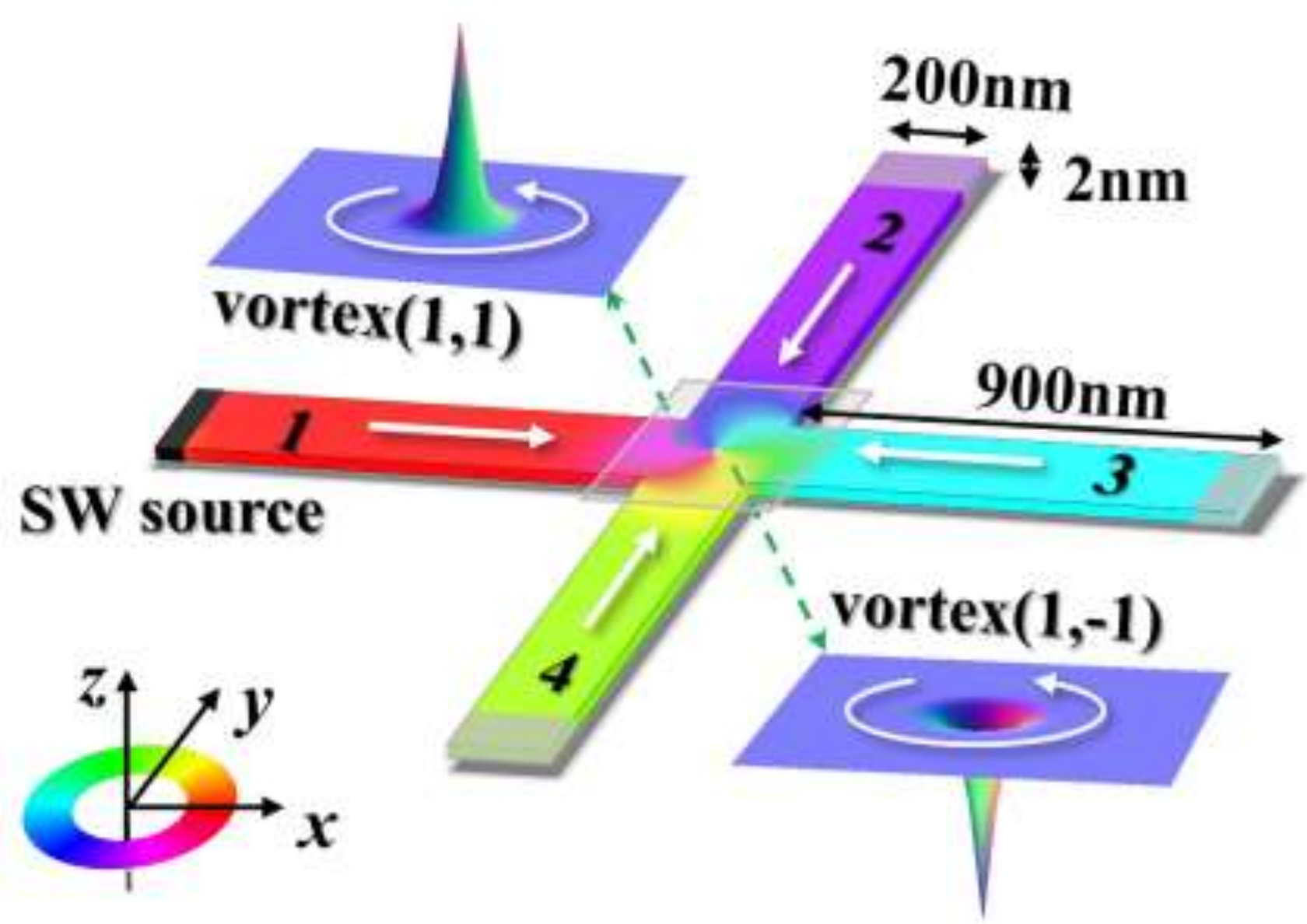

**Figure 1.** Geometry and dimension of the cross-shaped nanostructure in Cartesian coordinates. The color ring and white arrows indicate the in-plane magnetization distribution. The black and gray parts show the positions of wave source and high damping regions, respectively. The upper-left and lower-right insets represent the right- and left-handed magnetic vortex structure, respectively.

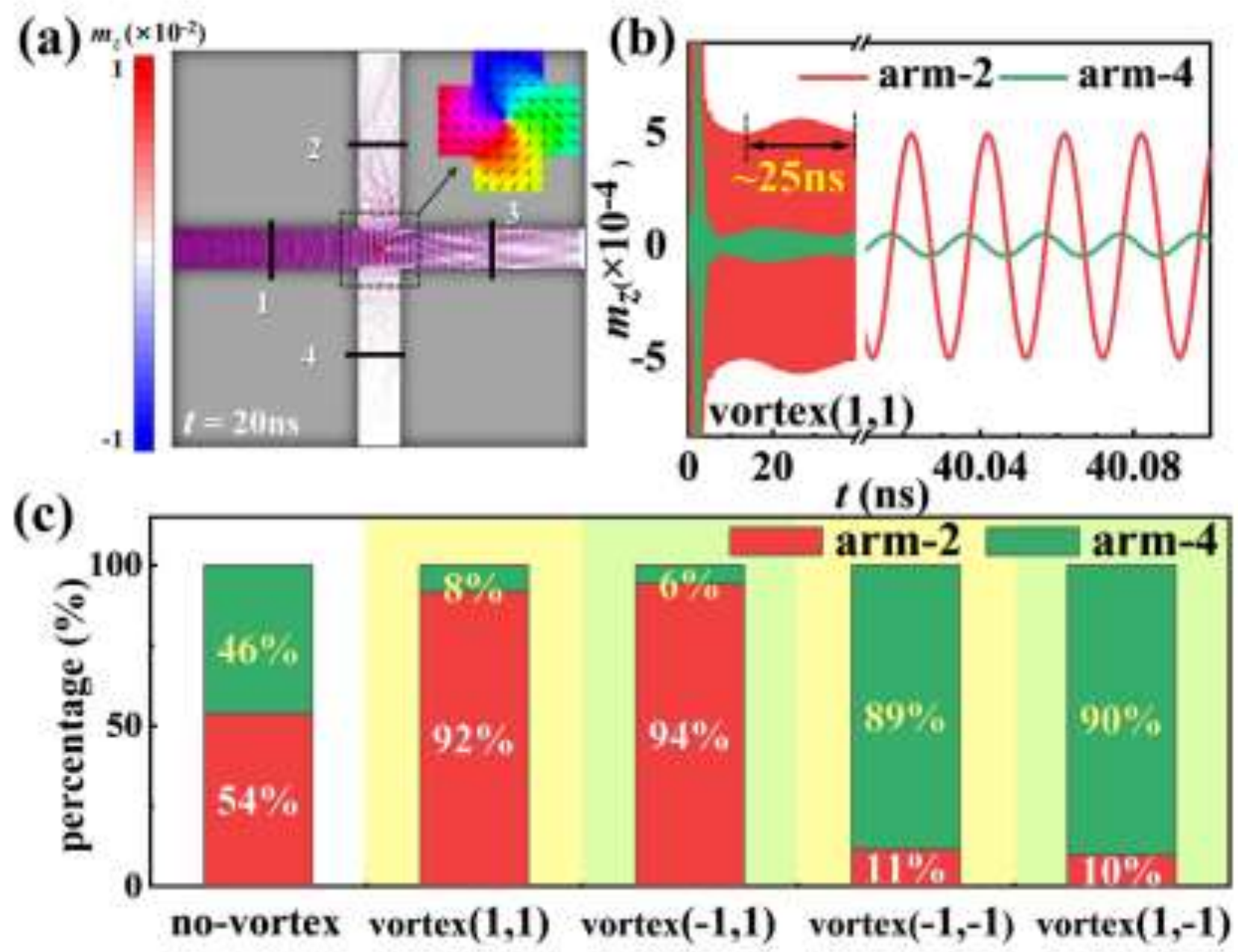

**Figure 2.** (a) Snapshots of the SW distribution in the nanostructure for the $\mathrm{vortex}(1,1)$ case at the $20\,\mathrm{ns}$. Color bar indicates the normalized $m_z$ value. Four black bars indicate the SW detecting positions, 550 nm far from the geometric center of the nanostructure. The inset show the spin structure of the $\mathrm{vortex}(1,1)$. (b) Time-varying $m_z$ signals collected at the detecting positions in arm-2 and arm-4 in the $\mathrm{vortex}(1,1)$ case. (c) The percentage stacked column of average amplitudes of SWs scattered into arm-2 and arm-4 in the cases of no-vortex, $\mathrm{vortex}(1,1)$, $\mathrm{vortex}(-1,1)$, $\mathrm{vortex}(-1,-1)$, and $\mathrm{vortex}(1,-1)$, respectively.

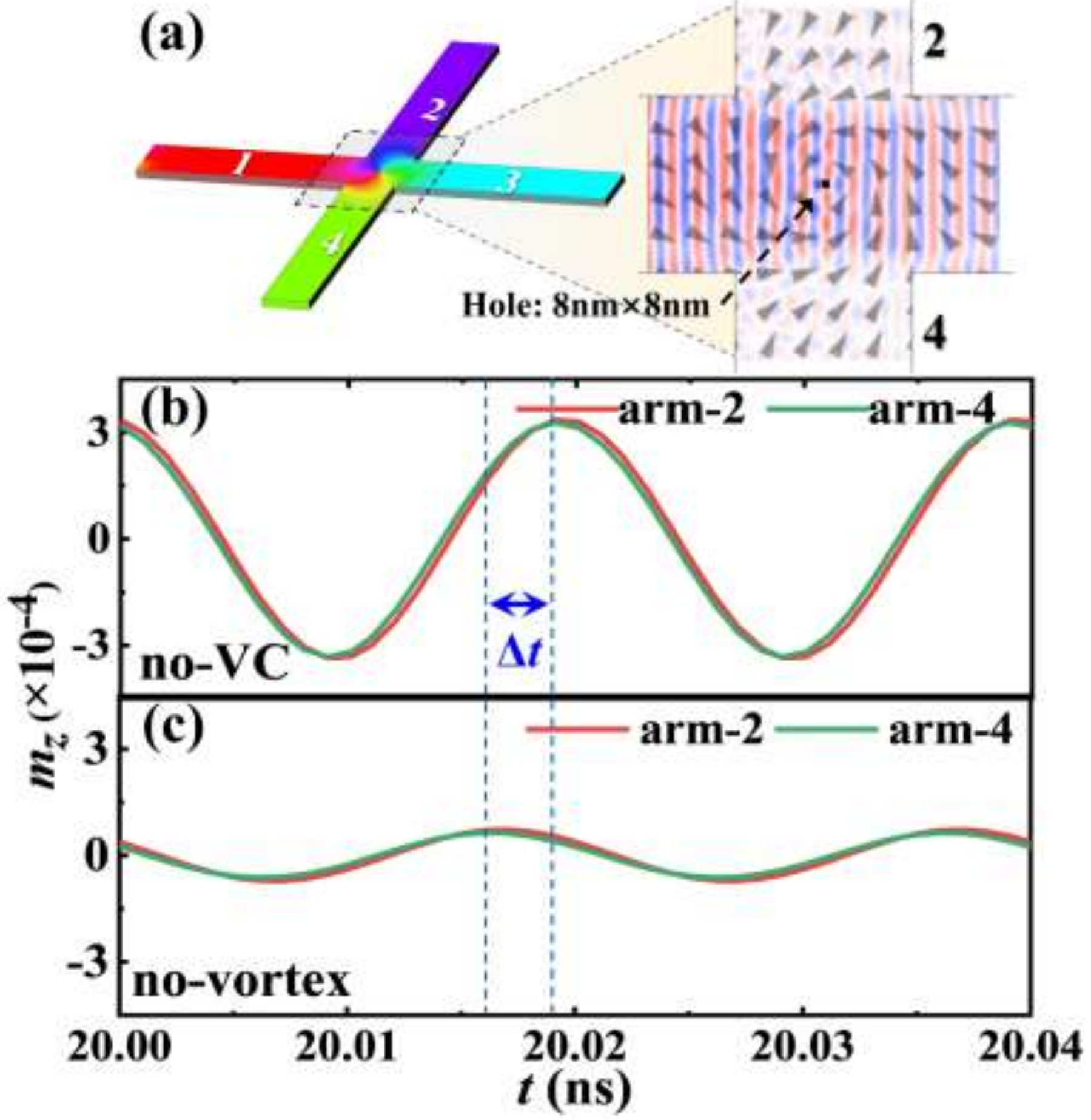

**Figure 3.** (a) Snapshots of SW distribution for the no-VC case at 20 ns after the SW excitation, the inset shows the magnetization distribution and SW distribution in the cross region. Time-varying $m_z$ signals collected at the detecting positions in arm-2 and arm-4, 550 nm far from the geometric center of the nanostructure, in the cases of (b) no-VC and (c) no-vortex.

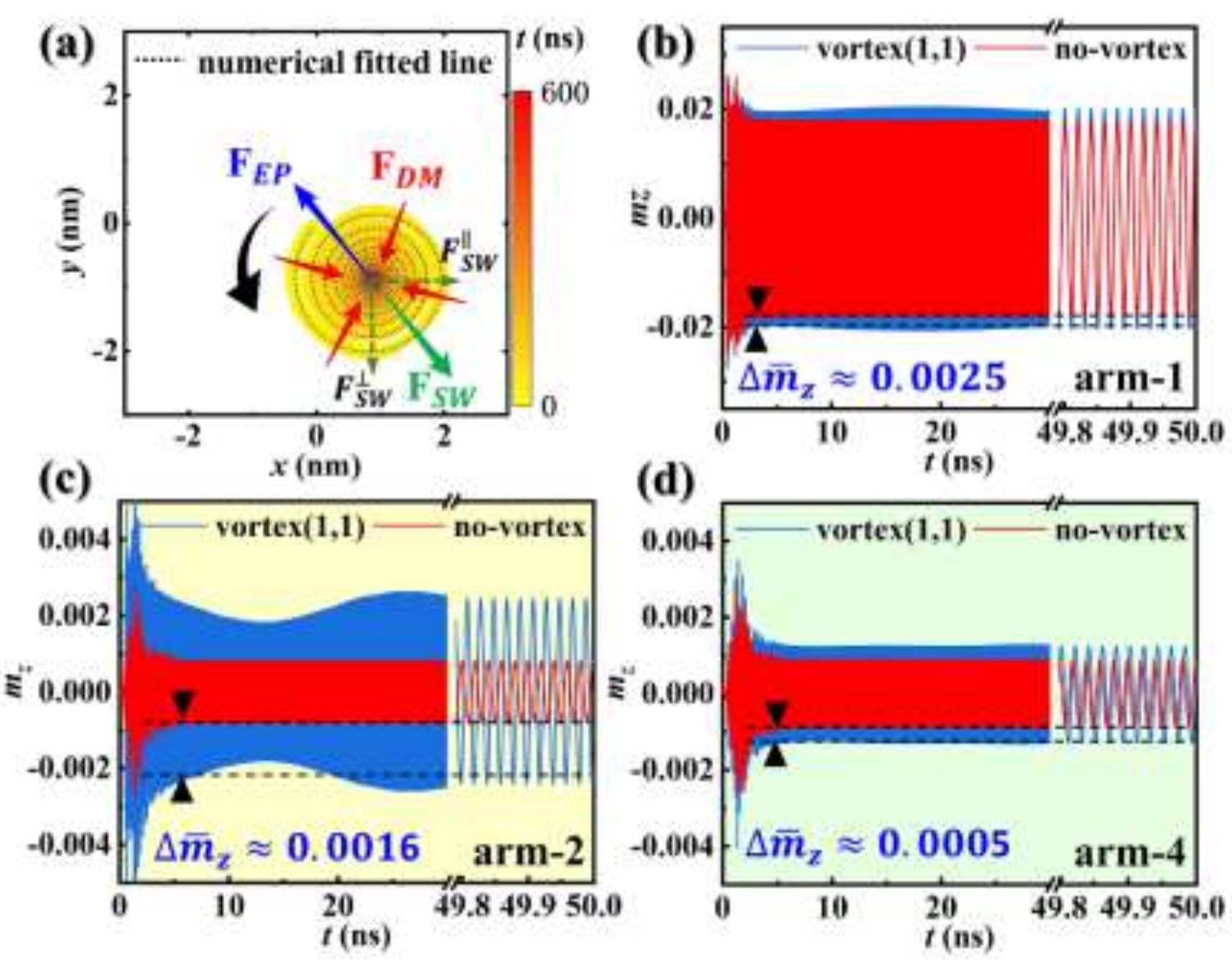

**Figure 4.** (a) The VC trajectory in the vortex(1,1) case during 600 ns, the black arrow represents the VC gyromotion direction and the red, blue and green arrows show the mechanical relationships between $F_{DM}$, $F_{EP}$ and $F_{SW}$. Color bar indicates the time variation and the dotted line represents the numerical fitting. (b) The time-varying $m_z$ signals collected at the detecting positions 160 nm far from the geometric center of the nanostructure are shown in (b) arm-1, (c) arm-2 and (d) arm-4, respectively, for the cases of vortex(1,1) and no-vortex.

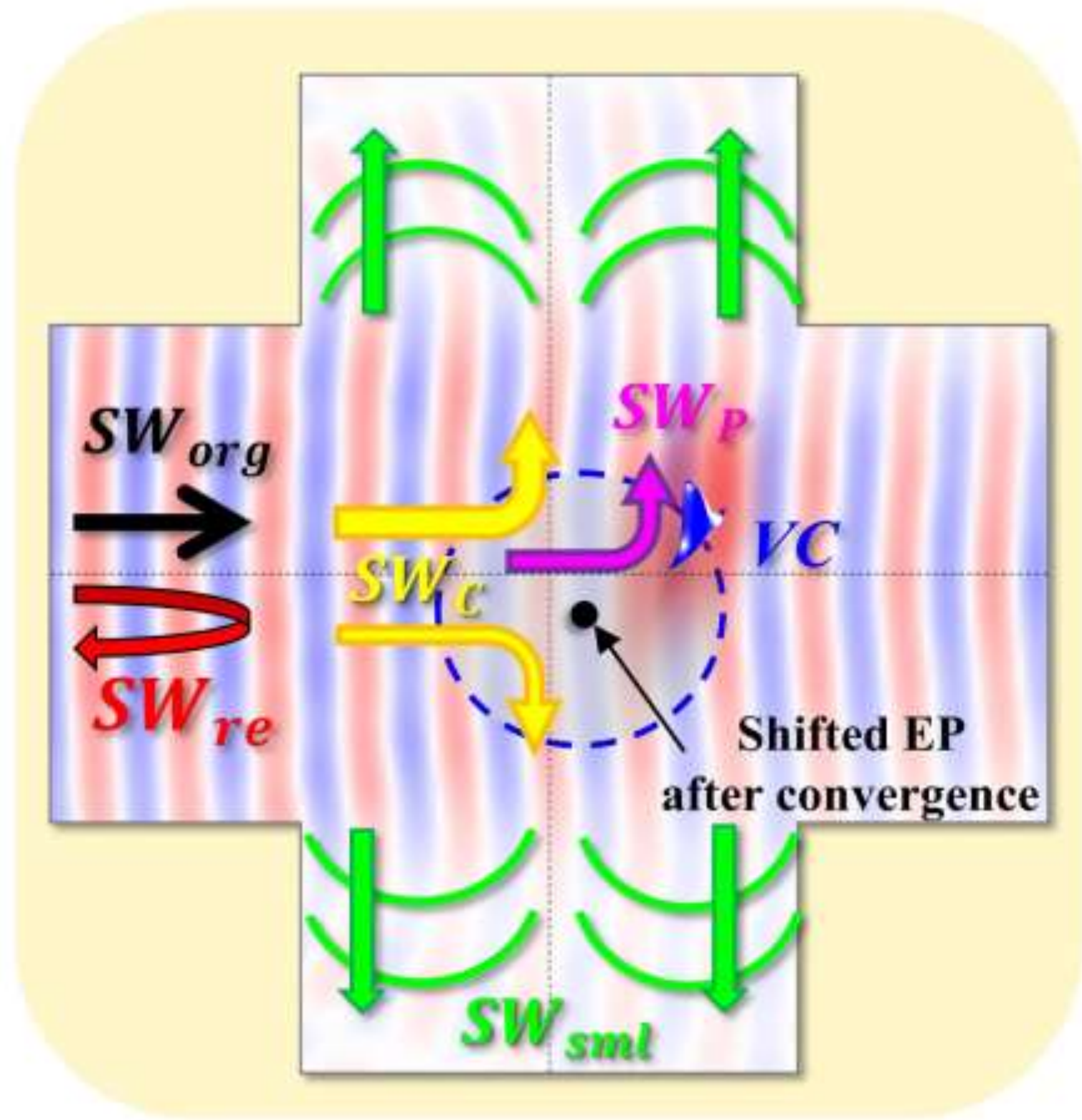

**Figure 5.** Schematic diagram of SW symmetric scattering mechanism for the SW propagating in the magnetic nonlinear system. The thick black arrow represents the $\mathrm{SW}_{org}$ propagating direction, the green arrows represent the geometric scatterings of $\mathrm{SW}_{sml}$, the yellow arrows represent the $\mathrm{SW}_{C}$ scatterings, the red curved arrow represents the $\mathrm{SW}_{re}$ reflection, and the purple arrow represents the skew scattering of $\mathrm{SW}_{P}$. The red and blue interference patterns represent the main propagation distribution of $\mathrm{SW}_{org}$.